\documentclass[10pt]{scrartcl}
\pdfoutput=1

\usepackage[T1]{fontenc}

\usepackage{amsmath,amsfonts}
\usepackage{booktabs}
\usepackage{color}

\usepackage{alltt}

\usepackage[margin=0.8cm, font=small, labelfont=bf]{caption}
\usepackage[list=true]{subcaption}
\DeclareCaptionSubType*[arabic]{figure}
\DeclareCaptionSubType*[arabic]{table}
\usepackage{etoolbox}

\let\theparentequation\theequation
\patchcmd{\theparentequation}{equation}{parentequation}{}{}

\newcommand*{\nextParentEquation}[1][]{
  \refstepcounter{parentequation}
  \setcounter{equation}{0}
  \ifx\\#1\\\relax\else\parentlabel{#1}\fi
}

\usepackage{jheppub}

\let\theparentequation\theequation
\patchcmd{\theparentequation}{equation}{parentequation}{}{}

\let\OLDthebibliography\thebibliography
\renewcommand\thebibliography[1]{
  \OLDthebibliography{#1}
  \setlength{\parskip}{.6pt}
  \setlength{\itemsep}{.6pt plus 0.3ex}
}

\newcommand{\unity}{\mathrm{1\mskip-4.5mu l}}
\def\zero{\mathbf{0}}
\newcommand{\IE}{i.e.\ }
\newcommand{\EG}{e.g.\ }
\newcommand{\I}{\mathrm{i}}
\newcommand{\E}{\mathrm{e}}

\newcommand{\FH}{\texttt{FeynHiggs}}%
\newcommand{\FA}{\texttt{FeynArts}}%
\newcommand{\FC}{\texttt{FormCalc}}%
\newcommand{\TC}{\texttt{TwoCalc}}%

\definecolor{xgrey}{gray}{.65}
\def\greyed#1{{\color{xgrey}#1}}

\definecolor{darkblue}{rgb}{0,0,.7}
\def\Code#1{\ensuremath{\texttt{#1}}}
\def\Var#1{\ensuremath{\texttt{\textsc{\color{darkblue}#1}}}}
\def\uscore{\symbol{95}}
\def\lbrac{\symbol{123}}
\def\rbrac{\symbol{125}}
\def\caret{\symbol{94}}

\usepackage{setspace}

\newcommand{\mathsym}[1]{{}}
\newcommand{\unicode}[1]{{}}

\BeforeTOCHead[toc]{{\pdfbookmark[1]{\contentsname}{toc}}}

\begin{document}


\thispagestyle{empty}

\def\thefootnote{\fnsymbol{footnote}}

\begin{flushright}
DESY 15-127\\
MPP-2015-157%
\end{flushright}

\vspace{2cm}

\begin{center}

{\large Implementation of the $\mathcal{O}(\alpha_t^2)$ MSSM Higgs-mass 
corrections in \FH}

\vspace{1cm}

Thomas Hahn$^1$\footnote{email: hahn@mpp.mpg.de}
and
Sebastian Pa{\ss}ehr$^2$\footnote{email: sebastian.passehr@desy.de}

\vspace*{.7cm}

\textsl{
$^1$Max-Planck-Institut f\"ur Physik \\
(Werner-Heisenberg-Institut), \\
F\"ohringer Ring 6, 
D--80805 M\"unchen, Germany
}
\\
\medskip
\textsl{
$^2$Deutsches Elektronen-Synchrotron DESY,\\
Notkestra{\ss}e 85, D--22607 Hamburg, Germany
}

\end{center}

\vspace*{2cm}

\begin{abstract}{}
  We describe the implementation of the two-loop Higgs-mass
  corrections of $\mathcal{O}(\alpha_t^2)$ in the complex MSSM in \FH.
  The program for the calculation is comprised of several scripts
  which flexibly use \FA{} and \FC{} together with other
  packages.  It is included in \FH{} and documented here in some 
  detail so that it can be re-used as a template for similar 
  calculations.
\end{abstract}

\def\thefootnote{\arabic{footnote}}
\setcounter{page}{0}
\setcounter{footnote}{0}

\newpage
\section{Introduction}

The Higgs-like boson discovered at
LHC~\cite{Aad:2012tfa,Chatrchyan:2012ufa} initiated great efforts to
pin it down as a particle responsible for electroweak symmetry
breaking.  Although the measured properties of this new boson are
consistent with the expectations for the Standard Model~(SM) Higgs
boson~\cite{Moriond2015}, a variety of beyond-SM interpretations is
still possible. One such possibility is the interpretation as a light
state within a richer spectrum of scalar particles in the
theoretically well motivated Minimal Supersymmetric Standard
Model~(MSSM). Its Higgs sector consists of two complex scalar doublets
leading to five physical Higgs bosons and three (pseudo-) Goldstone
bosons. The physical mass eigenstates at the tree level are the
neutral $CP$-even~$h$,~$H$, the $CP$-odd~$A$, and the
charged~$H^{\pm}$ bosons.  Their masses can be parameterized by the
charged (or $CP$-odd) Higgs-boson mass and the ratio of the two vacuum
expectation values, $\tan\beta = v_2/v_1$. $CP$-violation in the Higgs
sector is induced by complex parameters in other sectors of the MSSM
via loop corrections, leading to mixing between~$h$,~$H$, and~$A$ in
the mass
eigenstates~\cite{Pilaftsis:1998pe,Demir:1999hj,Pilaftsis:1999qt,Heinemeyer:2001qd}.

Since the masses of the neutral Higgs bosons are strongly affected by
loop contributions, a lot of work has been invested into higher-order
calculations of the mass spectrum from the SUSY parameters, in the
case of the real
MSSM~\cite{Heinemeyer:1998jw,Heinemeyer:1998np,Heinemeyer:1999be,Heinemeyer:2004xw,Borowka:2014wla,Degrassi:2014pfa,Borowka:2015ura,mhiggsFD3l,Zhang:1998bm,Espinosa:2000df,Brignole:2001jy,Casas:1994us,Degrassi:2002fi,Heinemeyer:2004gx,Allanach:2004rh,Martin:2001vx}
as well as for the MSSM with complex
parameters~(cMSSM)~\cite{Demir:1999hj,Choi:2000wz,Pilaftsis:1999qt,Heinemeyer:2001qd,Carena:2000yi,Frank:2006yh,Heinemeyer:2007aq,Hollik:2014wea,Hollik:2014bua}.
At one-loop order, the dominant contributions arise from the
Yukawa sector with the large top-Yukawa coupling~$h_t$,
or~\mbox{$\alpha_t=\left.h_t^2\middle/(4\pi)\right.$}. The class of
leading two-loop Yukawa-type corrections of~$\mathcal{O}(\alpha_t^2)$
has recently been calculated for the case of complex
parameters~\cite{Hollik:2014wea,Hollik:2014bua}. Together with the
full one-loop result~\cite{Frank:2006yh} and the
leading~$\mathcal{O}{\left(\alpha_{t}\alpha_{s}\right)}$
terms~\cite{Heinemeyer:2007aq} it is now available in the public
program
\FH{}~\cite{Heinemeyer:1998np,Degrassi:2002fi,Frank:2006yh,Heinemeyer:1998yj,Hahn:2009zz,Hahn:2010te}.

The $\mathcal{O}(\alpha_t^2)$ contributions in the cMSSM have been
evaluated in the Feynman-diagrammatic approach, extending the on-shell
renormalization scheme of Ref.~\cite{Frank:2006yh} to the two-loop
level. This ensures that the obtained analytical results for the
renormalized two-loop self-energies can consistently be incorporated
in \FH. For implementation in \FH{} the whole calculation of
the~$\mathcal{O}(\alpha_t^2)$ contributions was reworked and broken up
into several working steps. Together they provide a semi-automated
framework for a two-loop calculation in a model with non-trivial
renormalization. Although the code is specific to the evaluation of
the mentioned corrections, it can be utilized as a template for
similar calculations. The aim of this paper is to explain in detail
how the code works and what is done at each step, and where
process-specific adjustments are necessary.

A note on the software-engineering aspect is in order here, too: as many 
software packages (both in physics and elsewhere) also \FA{} and 
\FC{} have been designed to do a `complete' job.  That is, all the 
steps from the generation of Feynman diagrams to the numerical 
computation of a cross-section are executed from a single control 
program (\EG a single Mathematica session) -- that at least is how the 
demo programs insinuate usage.  There is nothing inherently wrong with 
this `monolithic' approach, it is maybe just not obvious how to extend 
it beyond the limitations of the package(s) used.

The suite of shell scripts and Mathematica packages we present here 
constitute a showcase of how to use \FA{} and \FC{} more flexibly, 
together with other packages.  The scripts provide the `outer hull' of a 
computation, in particular its compartmentalization, with coordination 
through a makefile.  The packages provide recurring tasks `library-like' 
in easily identifiable components.

The present code can be understood in a wider sense as a prototypical 
implementation of a new way of making flexible use of existing programs, 
even for very specific tasks, while retaining core functionality of the 
established packages.  It is in many aspects still a proof-of-concept 
realization since, for example, not all \FC{} routines can generically 
handle diagrams with more than one loop yet.  Likewise, a complete setup 
would include (standardized) converters to and from the more important 
available software packages, \EG for multiloop tensor reduction.  In 
that sense our code is not so much a demo program (to be blindly used) 
than a template (to be intelligently adapted).

The paper is organized as follows: Sect.~\ref{sec:HiggsSect} provides 
the theoretical foundation of the computation of the Higgs-mass 
corrections in the cMSSM.  The approximations used for the evaluation of 
the~$\mathcal{O}(\alpha_t^2)$ contributions to the Higgs masses are 
explained in Sect.~\ref{sec:appr}.  A detailed description of the 
implementation of the~$\mathcal{O}(\alpha_t^2)$ corrections in \FH{} is 
given in Sect.~\ref{sec:impl}.

\section{Masses of the Higgs bosons in the Complex MSSM\label{sec:HiggsSect}}

Since the calculation of the~$\mathcal{O}(\alpha_t^2)$ corrections has
already been presented in great detail
elsewhere~\cite{Hollik:2014wea,Hollik:2014bua} we shall recap only the
parts necessary for the implementation in \FH{} here.

In the basis of the tree-level Higgs states the loop-corrected mass 
matrices are given by~\cite{Frank:2006yh}
\begin{align}
\label{eq:massmatrices}
\mathbf{M}_0^2 &= \begin{pmatrix}
  m_h^2 - \hat\Sigma_{hh} & -\hat\Sigma_{hH} & -\hat\Sigma_{hA} & \greyed{-\hat\Sigma_{hG}} \\
  -\hat\Sigma_{hH} & m_H^2 - \hat\Sigma_{HH} & -\hat\Sigma_{HA} & \greyed{-\hat\Sigma_{HG}} \\
  -\hat\Sigma_{hA} & -\hat\Sigma_{HA} & m_A^2 - \hat\Sigma_{AA} & \greyed{-\hat\Sigma_{AG}} \\
  \greyed{-\hat\Sigma_{hG}} & \greyed{-\hat\Sigma_{HG}} &
  \greyed{-\hat\Sigma_{AG}} & \greyed{m_G^2 - \hat\Sigma_{GG}} \\
\end{pmatrix}, \\
\mathbf{M}_\pm^2 &= \begin{pmatrix}
  m_{H^\pm}^2 - \hat\Sigma_{H^-H^+} & \greyed{-\hat\Sigma_{H^-G^+}} \\
  \greyed{-\hat\Sigma_{G^-H^+}} & \greyed{m_{G^\pm}^2 - \hat\Sigma_{G^-G^+}}
\end{pmatrix}.
\end{align}
where the~$\hat\Sigma$ denote the renormalized self-energies. In the 
following we shall neglect the Goldstone matrix elements (greyed out 
above) since their numerical effect is tiny. Mixing with Goldstone bosons is 
taken into account inside the loop diagrams and for a consistent 
renormalization, of course. The physical masses are obtained as the 
real parts of the poles of the propagator matrix
\begin{align}
\Delta(p^2) = -\I\left[p^2\unity - \mathbf{M}^2\right]^{-1}.
\end{align}
This is tantamount to finding the zeroes of the determinant of 
$p^2\unity - \mathbf{M}^2$, which is core functionality of \FH.

For the cMSSM, the following contributions to 
Eq.~\eqref{eq:massmatrices} are available in \FH: the one-loop 
self-energies with full $p^2$~dependence~\cite{Frank:2006yh}, the 
leading momentum-independent two-loop~$\mathcal{O}(\alpha_s\alpha_t)$ 
self-energies~\cite{Heinemeyer:2007aq}, and now also 
the~$\mathcal{O}(\alpha_t^2)$~\cite{Hollik:2014bua} self-energies. The 
subleading two-loop $\mathcal{O}(\alpha_s\alpha_b)$, 
$\mathcal{O}(\alpha_t\alpha_b)$ parts are implemented only in the rMSSM 
yet, and are interpolated in the phases when complex parameters are 
chosen \cite{Frank:2006yh}.

For the latter corrections the following components have to be computed:
\begin{itemize}
\item The unrenormalized genuine two-loop self-energies $\Sigma_{hh}^{(2)}$,
  $\Sigma_{hH}^{(2)}$, $\Sigma_{hA}^{(2)}$, $\Sigma_{HH}^{(2)}$,
  $\Sigma_{HA}^{(2)}$, $\Sigma_{AA}^{(2)}$, $\Sigma_{H^+H^-}^{(2)}$ at
  $p^2 = 0$ in $\mathcal{O}(\alpha_t^2)$ approximation.
\item The one-loop diagrams with insertions of one-loop counterterms.
  For the~$\mathcal{O}{\left(\alpha_t^2\right)}$ corrections the
  couplings and masses of the colored sector and chargino--neutralino
  sector are affected by this subrenormalization.
\item The two-loop counterterms for these self-energies. The 
  counterterms, the imposed renormalization conditions, and all 
  renormalization constants are given in Ref.~\cite{Hollik:2014bua}.
\item The two-loop tadpoles $T_h^{(2)}$, $T_H^{(2)}$, $T_A^{(2)}$ in
  $\mathcal{O}(\alpha_t^2)$ approximation, which enter the two-loop 
  counterterms.
\end{itemize}

\section[Approximations in the \texorpdfstring{\boldmath{$\mathcal{O}{(\alpha_t^2)}$}}{order alpha top squared} contributions]{Approximations in the \boldmath{$\mathcal{O}{\scalebox{1.2}[1.3]{(}\alpha_t^2\scalebox{1.2}[1.3]{)}}$ contributions\label{sec:appr}}}
 
The dominant terms of the~$\mathcal{O}{(\alpha_t^2)}$ contributions
are enhanced by an additional factor~$m_t^2$. The following
approximations are applied to the renormalized two-loop self-energies
in Eq.~\eqref{eq:massmatrices} to yield only this leading part.

\subsection{Gaugeless limit}

The gauge couplings~$g_1$ and~$g_2$ are set to zero, and also the
strong coupling~$g_{\text{s}}$ is discarded.  Consequently, also the
gauge-boson masses~$M_W$ and~$M_Z$ vanish, while the weak mixing
angle~$\theta_{\text{w}}$ retains its original value.  These choices
cannot naively be substituted in the original MSSM model file of
\FA~\cite{Hahn:2001rv} where the couplings are not yet expressed in
terms of the top-Yukawa coupling $h_t =
e\,m_t/(\sqrt{2}\,s_{\beta}\,s_{\text{w}}\,M_W)$.  Furthermore, the
non-zero ratios $\delta M_W/M_W$ and $\delta M_Z/M_Z$ appear in the
renormalization of $h_t$, while the gauge-boson-mass renormalization
constants $\delta M_W$ and $\delta M_Z$ themselves vanish.  The
technical implementation of these issues is described in
Sect.~\ref{sec:impl}.

\subsection{Vanishing external momentum}

The external momentum is set to zero, \IE the integrals that appear in 
the two-loop self-energies and the renormalization constants are 
calculated at~\mbox{$p^2 = 0$}.  As an immediate consequence all appearing 
two-loop integrals are independent of~$p^2$, leading to vacuum diagrams 
that are known analytically.

\subsection{Bottom mass equal to zero}

To isolate the leading~$\mathcal{O}(\alpha_t^2)$ contributions only, the 
bottom-quark mass~$m_b$ has to be set to zero.

\subsection{Consequences of the approximations}

The approximations mentioned above modify the MSSM parameters that enter 
the calculation as follows:
\begin{itemize}
\item
The Higgs- and Goldstone-boson masses at the tree level become \mbox{$m_h^2 = m_G^2 = m_{G^\pm}^2 = 0$} and \mbox{$m_H^2 = m_A^2 = m_{H^\pm}^2$}.

\item
The Higgs-sector mixing angles $\alpha$ and $\beta$ are related through
$\alpha = \beta - \pi/2$.

\item
The stop masses are derived as the eigenvalues of
\begin{align}
  \begin{pmatrix}
    m_{\tilde t_{\text{L}}}^2 + m_t^2 & m_t X_t^* \\
    m_t X_t & m_{\tilde t_{\text{R}}}^2 + m_t^2
  \end{pmatrix},
\end{align}
with $X_t = A_t - \mu^*/t_\beta$, where the soft-SUSY-breaking parameters
$m_{\tilde{t}_{\text{L,R}}}^{2}$ and $A_t$ enter.

\item
Only one sbottom remains in the Feynman diagrams with $m_{\tilde b}^2 = 
m_{\tilde t_{\text{L}}}^{2}$.  The sbottom mixing matrix is set to the 
unity matrix.

\item
Of the neutralinos and charginos only the original 
Higgsinos~$\tilde{\chi}_{3,4}^0,\,\tilde{\chi}_2^\pm$ occur in the 
Feynman diagrams and their degenerate masses are equal to $|\mu|$.  The 
mixing matrices $\mathbf{N}$ for the neutralinos and $\mathbf{U}$, 
$\mathbf{V}$ for the charginos simplify to
\begin{align}
  \mathbf{N} &= 
  \begin{pmatrix}
    \begin{pmatrix}
      \E^{\frac{\I}{2} \phi_{M_{1}}} & 0 \\
      0 & \E^{\frac{\I}{2} \phi_{M_{2}}}
    \end{pmatrix} & \zero \\
    \zero  & \frac 1{\sqrt 2}\E^{\frac{\I}{2}\phi_{\mu}}
    \begin{pmatrix} 1 & -1\\ \I & \I \end{pmatrix} 
  \end{pmatrix},
&
  \mathbf{U} &= \begin{pmatrix}
    \E^{\I\phi_{M_{2}}} & 0 \\
    0 & \E^{\I\phi_\mu}
  \end{pmatrix},
&
  \mathbf{V} &= \unity \,.
\end{align}

\item Many couplings are simplified.  Technically this is taken care
  of by a modified version of the~\FA{} model file for the complex
  MSSM, generated as described in Sect.~\ref{sec:impl}.  Explicit results
  can also be found in Ref.~\cite{Hollik:2014bua}.
\end{itemize}

\section{Implementation in \FH{}\label{sec:impl}}

Our code for the calculation of the renormalized two-loop self-energies 
at $\mathcal{O}(\alpha_t^2)$ is based on several Mathematica Notebooks 
used in the former calculation~\cite{Hollik:2014bua}.  From the aspect 
of software development these Notebooks had various shortcomings: 
duplicate code (\EG gaugeless limit implemented multiply), parallel 
instructions poured all over, even the requirement to run with a 
particular Mathematica version -- signs that the use of different 
packages together with various special requirements was nontrivial and 
that the reorganization was helpful.  The new code is included in the 
\FH{} tarball and, while still specific to the corrections mentioned, is 
very adaptable and may serve as a template for similar calculations, 
which is the reason we describe it here in some detail.

The calculation has been divided into seven working steps implemented by 
the scripts described in the following sections.  Additionally, several 
packages that are required for specific purposes are accessed during 
each step. A makefile coordinates the entire calculation.  The total 
running time is about~15--20 minutes and the final output is a highly 
optimized and very compact Fortran code with a file size of about 350 
kBytes.  The result can directly be moved into the \FH{} source tree.  
The following diagram gives a flow chart of scripts and the external 
software packages used:

\begin{center}
\vspace{1ex}\includegraphics[width=.9\hsize]{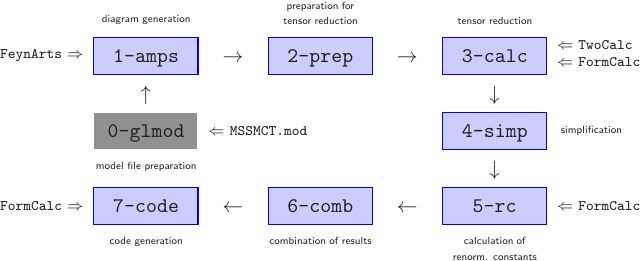}\vspace{1ex}
\end{center}

The advantages of this setup are twofold: First, the 
compartmentalization takes care of `concurrent' software packages (which 
cannot be loaded into the same Mathematica session, typically because of 
symbol conflicts).  Likewise, if an extension of this code required a 
different package for a particular step (say, tensor reduction), only 
this one step would need to be modified, and the additional package 
could be an arbitrary one, not just a Mathematica package, since it is 
invoked from a shell script. Second, the makefile sees to it that only 
the necessary parts of the calculation are redone in case of a change, 
which was a particular time-saver during the debugging phase.

Some details of the software setup:
\begin{itemize}
\item
The code described below is included in \FH{} 2.11.0 or later, in 
the \Code{gen/tlsp} tree.

\item The shell scripts (\Code{/bin/sh}) in the `\Code{scripts}' 
subdirectory perform the actual calculation.  They are executed directly 
from the command-line (\IE not loaded within Mathematica), usually by 
the makefile.  Internally they invoke the Mathematica Kernel through the 
shell's `here documents' as described in Ref.~\cite{Hahn:2006ig}.

\item The Mathematica programs in the `\Code{packages}' subdirectory 
provide functions for specific purposes and are read by the scripts.

\item As in the \FA{} model file for the MSSM~\cite{Hahn:2001rv}
we choose the particle labels
$$
h = \Code{h0},\quad
H = \Code{HH},\quad
A = \Code{A0},\quad
H^+ = \Code{Hp},\quad
H^- = \Code{Hm}.
$$

\item Each script is run from the command line with up to two arguments,
\EG
\begin{alltt}
   scripts/1-amps \Var{arg\(\sb1\)} \Var{arg\(\sb2\)}
\end{alltt}
\begin{tabbing}
where \=\Var{arg$_1$} =
\= \Code{h0h0}, \Code{h0HH}, \Code{h0A0}, \Code{HHHH},
   \Code{HHA0}, \Code{A0A0}, \Code{HmHp}
   (self-energies), \\
\>\> \Code{h0}, \Code{HH}, \Code{A0}
   (tadpoles), \\[.5ex]
\> \Var{arg$_2$} =
\> \Code{0}~~\= for virtual two-loop diagrams, \\
\>\> \Code{1} \> for one-loop diagrams with one-loop counterterms.
\end{tabbing}
That is, each self-energy/tadpole is computed and stored individually.  
Arguments are checked upon entry and meaningless arguments are omitted 
(\EG no \Var{arg$_2$} after combination of virtual and counterterm 
diagrams).

\item Symbolic expressions are generally output to the `\Code{m}' 
directory, Fortran code to the `\Code{f}' directory.

\item Input and output files are defined in statements `\Code{in=...}' 
and `\Code{out=...}' in the first few lines of each script.  In lieu of 
\textsl{in vivo} debugging, \EG setting breakpoints or inspecting 
variables, which is not easily possible with a shell script we have set 
up detailed log files.  A script's session dialog is written to the 
output file with \Code{.log.gz} appended.  For example, 
\Code{m/h0h0/6-comb.log.gz} is the log of the run that made 
\Code{m/h0h0/6-comb}.

\end{itemize}

\subsection{Model-file preparation, Gaugeless Limit (\Code{0-glmod})}
\label{sec:mssmctgl}

Our calculation uses the \FA{} model file for the MSSM with 
counterterms~\cite{Fritzsche:2013fta}.  Since the calculation is carried 
out in the gaugeless limit (cf.\ Sect.~\ref{sec:appr}), it speeds up 
calculations to prepare a variant of the model file where this limit is 
taken already at the level of the Feynman rules.  At this stage we also 
introduce $h_t$ (where $h_t^2 = 4\pi\alpha_t$) and substitute $A_t$ by 
$X_t + \mu^*/t_\beta$.  This is carried out by the \Code{0-glmod} script:
\begin{itemize}
\item Load the \Code{MSSMCT.mod} model file.
\item Modify/simplify couplings and remove those that become zero.
\item Write out an \Code{MSSMCTgl.mod} model file.
\end{itemize}

The original \Code{MSSMCT.mod} model file contains both the counterterm 
vertices and the definitions of the (one-loop) renormalization 
constants.  We moved the latter to \Code{model/MSSMCT.rc1} to make 
modification of the Feynman rules easier.

For each step we shall summarize usage as follows:
\begin{itemize}
\item Command line: \Code{model/0-glmod}

\item Main packages used:
\Code{packages/Gaugeless.m}, 
\Code{packages/XtSimplify.m}

\item Input: \Code{model/MSSMCT.mod.in}

\item Output: \Code{model/MSSMCT.mod}, \Code{model/MSSMCTgl.mod}
\end{itemize}

\subsection{Step 1: Diagram generation (\Code{1-amps})}

Diagrams are generated with \FA{}~\cite{Hahn:2000kx} using the
\Code{MSSMCTgl.mod} model file prepared above. Customized wrapper
functions for the \FA{} functions are provided by
\Code{packages/FASettings.m}.

They ensure that all parts of the calculation consistently use the same 
\FA\ settings.  Also diagram selection is simplified by filters defined 
in \Code{FASettings.m}.  Defining a diagram filter which works for an 
arbitrary topology is not entirely straightforward at the two-loop level 
but since only a few topologies contribute we have chosen a per-topology 
approach where particle choices can be made for each propagator.

\begin{figure}
\centerline{\includegraphics[width=.65\hsize]{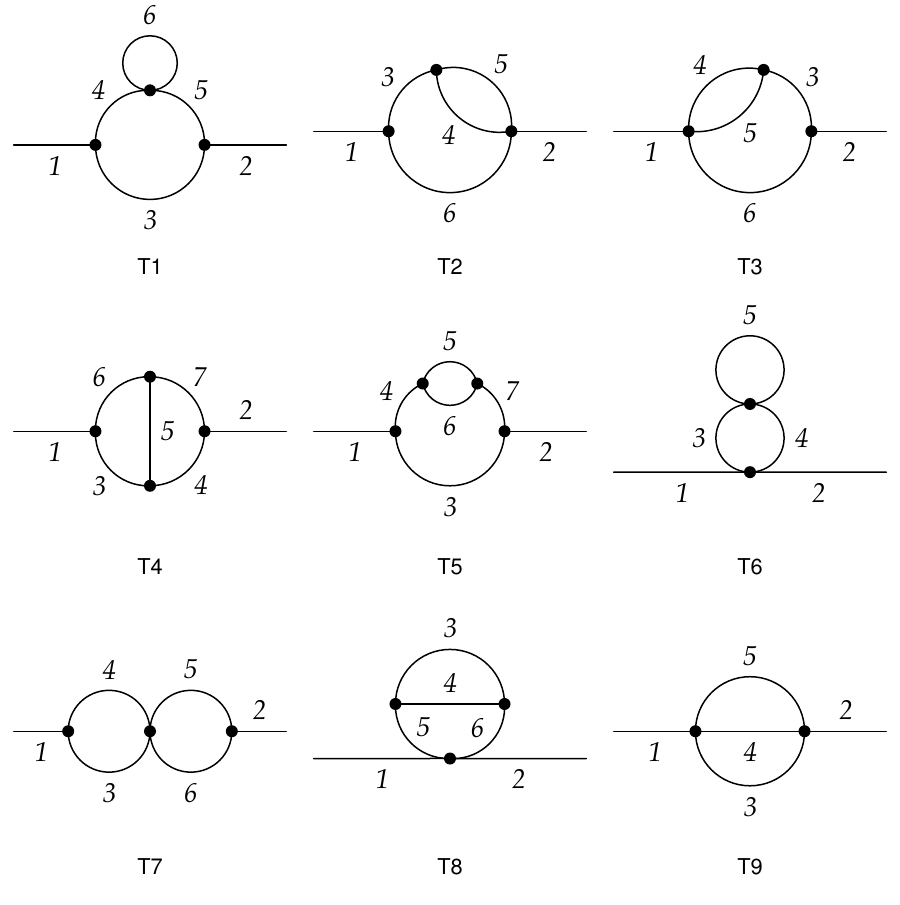}}
\caption{\label{fig:top2L}The 1PI two-point two-loop topologies with
  propagator numbers.}
\end{figure}

For example, the two-loop self-energy diagrams for the neutral Higgs
bosons are picked with the following statement:
\begin{verbatim}
   sel[0][S[_] -> S[_]] = {
     t[3] && htb[6],
     t[3] && tb[6],
     t[3] && tb[6],
     t[3] && t[4] && htb[5],
     t[3] && htb[5|6],
     t[3] && htb[5],
     t[3] && t[5],
     t[5] && ht[3|4],
     t[3|4|5] && ht[3|4|5] }
\end{verbatim}
The nine elements of the list correspond to the nine two-loop topologies 
of Fig.~\ref{fig:top2L}.  For example, the entry \Code{t[3]} mandates 
that propagator \#3 must carry a top or stop.  Likewise, \Code{htb[5|6]} 
enforces a Higgs/Higgsino, top/stop, or bottom/sbottom on propagators 
\#5 and/or \#6.

\begin{itemize}
\item Command line: \Code{scripts/1-amps} \Var{se} \Var{cto}

\Var{se} is one of the self-energy/tadpole names given above
(\Code{h0h0}, \Code{HHHH}, etc.)\ and \Var{cto} is the counterterm
order, \IE 0 for the virtual two-loop diagrams and 1 for the
one-loop plus counterterm diagrams.

\item Main packages used:
\Code{packages/FASettings.m}, 
\Code{packages/Gaugeless.m}

\item Input: (none)

\item Output:
\Code{m/\Var{se}/1-amps.\Var{cto}} (amplitude),
\Code{m/\Var{se}/1-amps.\Var{cto}.ps.gz} (diagrams)
\end{itemize}

\subsection{Step 2: Preparation for Tensor Reduction (\Code{2-prep})}

The tensor reduction is traditionally the step which increases the 
number of terms in the amplitude most, therefore we inserted Step 2 
before and Step 4 after the tensor reduction to keep the number of terms 
as small as possible at all times.

Several simplifications are carried out in Step~2: Firstly, the 
limit~\mbox{$p^2\to 0$} is applied. This limit is part of the 
approximation employed here and moreover admits closed-form expressions 
(logs and dilogs) for the two-loop integrals.

Secondly, simplifications are performed in particular on the ubiquitous 
sfermion mixing matrix elements\footnote{%
  The \Code{USf} actually carry two more indices for sfermion type and
  generation: \Code{USf[$i$,$j$,$t$,$g$]}, which we suppress here.}
$U_{ij}\equiv\Code{USf[$i$,$j$]}$ $(i, j = 1, 2)$.  For $2\times 2$ 
matrices the unitarity relations $U U^\dagger = U^\dagger U = \unity$ 
are of course easily written out, \EG $U_{11} U_{11}^* + U_{12} U_{12}^* 
= 1$; the problem is that Mathematica by itself rarely arranges 
expressions in just the way that any of these sums can be substituted 
directly.  This can be improved by adding definitions for single 
summands, \EG $|U_{22}|^2 = |U_{11}|^2$.  Because these apply 
\emph{while} Mathematica ponders the simplification strategy they 
increase the incentive for \Code{Simplify} to choose the 
unitarity-simplified version.  Mathematica does not allow to assign 
directly to the product of two matrix elements, however (technically the 
assignment is too `deep' for Mathematica's upvalues\footnote{%
	Mathematica associates a definition like \Code{A*B\caret 2 = 2}
	(internal notation: \Code{Times[A,\,B\caret 2] = 2}) with the
	multiplication function \Code{Times}.  For efficiency and also
	because \Code{Times} is a protected function, Mathematica 
	alternately allows the association with \Code{A}, called an 
	upvalue of \Code{A}.  The association cannot be made with 
	\Code{B}, however, because \Code{B} appears squared and hence 
	the assignment would be more than one level `above' \Code{B}.}),
hence we must introduce intermediate symbols:
\begin{align}
\Code{UCSf[$i$,$j$]} &= \Code{USf[$i$,$j$]}\,\Code{USf[$i$,$j$]}^*, \\
\Code{UCSf[$i$,3]} &= \Code{USf[$i$,1]}\,\Code{USf[$i$,2]}^*, &
\Code{UCSf[3,3]} &= \Code{USf[1,1]}\,\Code{USf[2,2]}^*, \\
\Code{UCSf[3,$j$]} &= \Code{USf[1,$j$]}\,\Code{USf[2,$j$]}^*, &
\Code{UCSf[3,4]} &= \Code{USf[1,2]}\,\Code{USf[2,1]}^*.
\end{align}
The unitarity relations can now easily be formulated through the 
\Code{UCSf}, \EG \Code{UCSf[2,2]} = \Code{UCSf[1,1]}.  The package
\Code{USfSimplify.m} contains the complete set of such rules and the 
function \Code{USfSimplify} which encodes the above procedure.

After applying unitarity it is observed that almost all \Code{UCSf} 
in the amplitude appear in the form of only four linear 
combinations:
\begin{align}
\Code{U2s1[$x$]}
  &= (\Code{UCSf[1,3]}\,x^* + \Code{UCSf[1,3]}^*\,x)/2\,, 
\label{eq:U2s1} \\
\Code{U2s2[$x$]} 
  &= (\Code{UCSf[1,3]}\,x^* - \Code{UCSf[1,3]}^*\,x)/(2\I)\,,
\label{eq:U2s2} \\
\Code{U2c1[$x$]}
  &= (\Code{UCSf[3,3]}\,x^* + \Code{UCSf[3,4]}^*\,x)/2\,,
\label{eq:U2c1} \\
\Code{U2c2[$x$]}
  &= (\Code{UCSf[3,3]}\,x^* - \Code{UCSf[3,4]}^*\,x)/(2\I)\,.
\label{eq:U2c2}
\end{align}
Substituting these combinations indeed shortens the amplitude by about 
1/3.  Again, this choice is empirical and may well be specific to the 
corrections computed here.  All relations for substitution and 
simplification are contained in \Code{packages/U2Simplify.m}.  Note in 
particular that, despite the suggestive notation, Eqs.~\eqref{eq:U2c1} 
and \eqref{eq:U2c2} are complex quantities since the two summands are 
not conjugates of each other.

\begin{itemize}
\item Command line: \Code{scripts/2-prep} \Var{se} \Var{cto}

\Var{se} and \Var{cto} as in Step~1.

\item Main packages used:
\Code{packages/USfSimplify.m}, 
\Code{packages/U2Simplify.m}, \\
\Code{packages/SimplificationDefinitions.m}

\item Input: \Code{m/\Var{se}/1-amps.\Var{cto}}

\item Output: \Code{m/\Var{se}/2-prep.\Var{cto}}
\end{itemize}

\subsection{Step 3: Tensor Reduction (\Code{3-calc})}

The tensor reduction of the loop integrals is done through 
relatively straightforward application of the packages 
\TC{}~\cite{Weiglein:1993hd,Weiglein:1995qs} (for the virtual two-loop 
diagrams, \Var{cto} = 0) and \FC{}~\cite{Hahn:1998yk} (for the one-loop 
plus counterterm diagrams, \Var{cto} = 1).

Noteworthy here is perhaps that \TC{} and \FC{} cannot be loaded into 
one Mathematica session due to symbol conflicts, but this is easily 
accomodated in the shell-script setup, \IE the shell script invokes not 
one but two different Mathematica sessions depending on the value of 
\Var{cto}.

After the reduction, scaleless integrals (which vanish in dimensional 
regularization) are removed using \Code{packages/UseSimplePackage.m}.

\begin{itemize}
\item Command line: \Code{scripts/3-calc} \Var{se} \Var{cto}

\Var{se} and \Var{cto} as in Step~1.

\item Main packages used:
\TC{}, \FC{},
\Code{packages/FCSettings.m}, \\
\Code{packages/UseSimplePackage.m}

\item Input: \Code{m/\Var{se}/2-prep.\Var{cto}}

\item Output: \Code{m/\Var{se}/3-calc.\Var{cto}}
\end{itemize}

\subsection{Step 4: Simplification after Tensor Reduction (\Code{4-simp})}

Step~4 is the counterpart of Step~2 after tensor reduction and reduces 
the size of each individual amplitude before they are combined in 
Step~6.

The key to decent performance comes from a `tag' function\footnote{%
	The \Code{DiagMark} function was originally developed for
	Refs.~\cite{Heinemeyer:2003dq,Heinemeyer:2004yq}.}
\Code{DiagMark[$m_i$]} which the \FA{} wrapper functions of Step~1 
(\Code{packages/FASettings.m}) inserted for each diagram, where $m_i$ 
are the masses that run in the loop.  This \Code{DiagMark} function 
induces a partitioning, \IE the entire amplitude can be written as the 
sum of pieces, each of which is multiplied by a different 
\Code{DiagMark}.  Typically only few simplifications can be made across 
these pieces, hence one can restrict application of a simplification 
function \Var{simp} to each piece, as in
\begin{quote}
   \Code{Collect[\Var{amp}, \uscore DiagMark, \Var{simp}]}
\end{quote}
Mathematica's \Code{Collect} function reorders \Var{amp} as a linear 
combination of \Code{DiagMark} instances and applies \Var{simp} to each 
coefficient, which is of course much faster than applying it to the 
entire expression.

\begin{itemize}
\item Command line: \Code{scripts/4-simp} \Var{se} \Var{cto}

\Var{se} and \Var{cto} as in Step~1.

\item Main packages used:
\Code{packages/SimplificationDefinitions.m},
\Code{packages/FCSettings.m}

\item Input: \Code{m/\Var{se}/3-calc.\Var{cto}}

\item Output: \Code{m/\Var{se}/4-simp.\Var{cto}}
\end{itemize}

\subsection{Step 5: Calculation of the Renormalization Constants 
(\Code{5-rc})}

The one-loop Renormalization Constants (RCs) are defined in 
\Code{MSSMCT.rc1} (see Sect.~\ref{sec:mssmctgl}), where the $\delta 
M_V^2$ ($V = W, Z$) require special attention, however.  As already 
pointed out in Sect.~\ref{sec:appr}, they are proportional to $M_V^2$ 
and would naively vanish in the gaugeless limit, while $\delta 
M_V^2/M_V^2$ must not vanish.  Technically, we accomplish this by 
explicitly writing out the proportionality to $M_V^2$: 
\mbox{$\Code{dMVsq1} \to \Code{dMVsq1MV2*MV2}$}, such that a possible 
\Code{MV2} in the denominator of the coefficient of \Code{dMVsq1} can 
cancel.

For the two-loop RCs of the cMSSM no complete set of definitions is
available as a model file yet.\footnote{%
  A complete list for the counterterms of the Higgs sector in the
  complex MSSM can be found in Ref.~\cite{Hollik:2014bua}.}
We added the necessary ones in the approximation relevant here in
\Code{model/MSSMCT.rc2}.  For better extendability we tried to
preserve the mathematical notation as much as possible, \EG the
two-loop wavefunction renormalization is defined via (the mass
counterterm is defined separately)
\begin{verbatim}
   dMHmat[Msq_, dZ1_, dZ2_, dM1_] := 
     1/2 (Msq.dZ2 + ConjugateTranspose[dZ2].Msq) +
     1/2 (ConjugateTranspose[dZ1].dM1 + ConjugateTranspose[dM1].dZ1) +
     1/4 ConjugateTranspose[dZ1].Msq.dZ1
\end{verbatim}
which preserves the matrix notation of Eq.~(2.30) of \cite{Hollik:2014bua}.

The RCs are further expanded in the dimension parameter~$\varepsilon =
(D - 4)/2$.  For example, the one-loop top-mass RC, \Code{dMf1[3,3]},
is decomposed as
\begin{verbatim}
   dMf1[3,3] = RC[-1, dMf1[-1,3,3]] + RC[0, dMf1[0,3,3]]
\end{verbatim}
where the two terms on the r.h.s.\ represent the coefficients of 
$\varepsilon^{-1}$ and $\varepsilon^0$, respectively.  The loop 
integrals are expanded with the \Code{ExpandDel} function and powers of 
$\varepsilon$ are collected with \Code{DelSeries}, both defined in 
\Code{packages/ExpandDel.m}.  The notation allows to easily extract the 
terms actually needed in the final result in Step~6.

The $\mathcal{O}(\varepsilon^1)$ parts of the one-loop integrals often 
cancel in the final result.  This cancellation should (and in the 
present case has been) checked, yet it needs some effort to see the 
coefficients of the $\varepsilon^1$-terms actually vanish.  Once this is 
established, however, it is useful to set the $\varepsilon^1$-terms to 
zero manually, which is faster and gives a more compact result. In 
\Code{packages/ExpandDel.m} this is done through rules defined in 
\Code{AssumeVanishingDel}.  To re-enable checking, set 
\Code{AssumeVanishingDel = \lbrac\rbrac}.

The actual calculation of the RCs is carried out with \FC{}, for each 
self-energy and tadpole.  So as not to re-compute RCs many times, they 
are cached in the \Code{m/rc} subdirectory.  Take care that this cache 
is not controlled by the makefile.

\begin{itemize}
\item Command line: \Code{scripts/5-rc} \Var{se}

\Var{se} as in Step~1.

\item Main packages used:
\Code{packages/RenConst.m},
\Code{packages/ExpandDel.m}

\item Input: \Code{m/\Var{se}/4-simp.1}

\item Output: \Code{m/\Var{se}/5-rc}, \Code{m/rc/*}
\end{itemize}

\subsection{Step 6: Combining Results (\Code{6-comb})}

The final analytical step for each self-energy/tadpole joins the
amplitudes with the renormalization constants and, unless a debug flag
is set, retains only the coefficient of $\varepsilon^0$, \IE the
finite part. It also applies a (mostly empirical) blend of
simplification functions to get the compact expression for code
generation.

Two of these simplifications merit a closer look, for they affect also 
the numerical precision of the final result.  The difference and sum
of the sfermion mass squares not only appear frequently in the 
amplitude, they are also computed as a by-product of diagonalization 
anyway, with effectively no loss in precision even for degenerate 
sfermion masses.  We thus substitute
\begin{quote}
   \Code{MSf2[3,$t$,$g$] = MSf2[2,$t$,$g$] - MSf2[1,$t$,$g$]}, \\
   \Code{MSf2[4,$t$,$g$] = MSf2[1,$t$,$g$] + MSf2[2,$t$,$g$]}.
\end{quote}
Considering that in the gaugeless limit with vanishing bottom mass the
remaining sbottom mass square is equal to the stop-sector breaking
parameter, $m_{\tilde b}^2 = m_{\tilde t_{\text{L}}}^{2}$, the
difference of stop and sbottom mass squares can be computed with
better precision, too, which is particularly relevant for the $H^\pm$
self-energy where this difference appears in denominators.  Hence we
also introduce
\begin{quote}
   \Code{MSq2Diff[$i$,$j$] = MSf2[$i$,4,3] - MSf2[$j$,3,3]}.
\end{quote}

\begin{itemize}
\item Command line: \Code{scripts/6-comb} \Var{se}

\Var{se} as in Step~1.

\item Main packages used:
\Code{packages/FinalSimp.m},
\Code{packages/ExpandDel.m}

\item Input:
\Code{m/\Var{se}/4-simp.0},
\Code{m/\Var{se}/4-simp.1},
\Code{m/rc/*}

\item Output: \Code{m/\Var{se}/6-comb}
\end{itemize}

\subsection{Step 7: Generating Code (\Code{7-code})}

The last step writes out optimized Fortran code using~\FC{}'s 
code-generation functions \cite{Hahn:2010zi}.  Self-energies and 
tadpoles are arranged in three groups, such that \FH{} can also perform 
a partial evaluation depending on its flags (\EG if $2\times 2$ mixing 
only is requested):
\begin{quote}
1. \Code{A0A0}, \Code{HmHp}, \\
2. \Code{h0h0}, \Code{HHHH}, \Code{h0HH}, \Code{h0}, \Code{HH}, \\
3. \Code{h0A0}, \Code{HHA0}, \Code{A0}.
\end{quote}
Abbreviations are introduced for both loop integrals and common 
subexpressions, which shortens the entire expression significantly.  
Some variables need to be disambiguated, for example the sfermion masses 
in the gaugeless limit must not use the same identifiers as other 
sfermion masses in \FH.

Lastly, the handwritten part of the code from the `\Code{static}' 
subdirectory is copied to the Fortran output directory `\Code{f}', which 
is then complete and replaces the entire \Code{src/TwoLoop/TLsp} branch 
of the \FH{} source tree.

\begin{itemize}
\item Command line: \Code{scripts/7-code}

\item Main packages used:
\FC{}, \Code{packages/FCSettings.m}

\item Input: \Code{m/*/6-comb}, \Code{static/*}

\item Output: \Code{f/*}
\end{itemize}

\section{Conclusions}

We have described the implementation of the~$\mathcal{O}(\alpha_t^2)$ 
Higgs-mass corrections for the complex MSSM in \FH{}.  The original 
calculation of Ref.~\cite{Hollik:2014bua} was restructured entirely for 
this purpose and the new code is included in the \FH{} source code,
available at \Code{http://feynhiggs.de}.  The calculation is split 
into seven tasks which were explained in detail.

While each step is specific to the corrections computed here the code 
can be used as a template due to its modular structure and adapted 
with little effort to similar calculations.  The scripting framework 
we presented can also be considered a showcase of how to flexibly use
\FA{} and \FC{} together with other packages to generalize
applicability beyond one-loop level.

As with all practical recipes, the proof is in the pudding.  Our 
description herein should suffice to get an overview of the code and its 
structure.  What cannot be adequately conveyed in writing is that 
significant effort has been spent to make the code readable and 
`obvious'.

\section*{Acknowledgments}
\sloppy{We thank Dominik St\"ockinger for the \Code{DiagMark} function 
and other simplification recipes and Sven Heinemeyer for constructive 
comments on the manuscript. This work has been supported by the 
Collaborative Research Center SFB676 of the DFG, ``Particles, Strings 
and the early Universe.''}

\pdfbookmark[1]{References}{refs}

\end{document}